\documentclass[epj]{svjour}
\usepackage{graphics}
\begin{document}
\title{Majorana mode stacking, robustness and size effect in cylindrical nanowires}
\author{
Javier Osca\inst{1}\thanks{\emph{e-mail:} javier@ifisc.csic-uib.es} 
\and Rosa L\'opez\inst{1}
\and Lloren\c{c} Serra\inst{1,2} 
%
}                     
%
%
\institute{
Institut de F\'{\i}sica Interdisciplin\`aria i de Sistemes Complexos
IFISC (CSIC-UIB), E-07122 Palma de Mallorca, Spain 
\and 
Departament de F\'{\i}sica,
Universitat de les Illes Balears, E-07122 Palma de Mallorca, Spain}
\date{Received: date / Revised version: date}
%
\abstract{
We discuss the robustness of Majorana edge modes in a finite quantum nanowire 
of cylindrical shape.
The nanowire is modeled as a bidimensional cylindrical shell
of semiconductor material with proximity-induced superconductivity
and an intrinsic Rashba spin-orbit interaction. The latter is characterized by effective 
electric and magnetic fields in transverse direction of the nanowire. 
An applied external magnetic field pointing in an arbitrary orientation is also assumed. 
The numerical diagonalization of the Hamiltonian allows us to study the spectrum of the nanowire for different experimental configurations. The Majorana modes prove robust against tilting of the magnetic field away from the cylinder longitudinal axis, if the tilt direction is perpendicular to the effective spin-orbit magnetic field, but fragile otherwise. On the other hand, we find an increasing number of Majorana modes in the same cylinder edge for increasing values of the nanowire radius. We refer to this phenomenon as ``stacking effect" and it occurs due to the orthogonality between Majorana mode wave functions. In this manner, different Majoranas take complementary positions on the nanowire surface.
} 
\maketitle
\section{Introduction}
\label{intro}
The discovery of topological states of matter was a major milestone in the condensed matter field \cite{Kitaev,Wilczeck,Brouwer}.
These states behave as localized non abelian anyons, meaning that a nontrivial phase modification is obtained after the interchange
of a pair of them \cite{Pachos}. It has been argued that in addition to new yet to discover physics, their peculiar statistics has the potential of making these states the basic units for quantum processing and opens the possibility of achieving topological quantum
computation \cite{Nayak}. The nonlocal properties of these topological quasiparticles gives them a certain degree of immunity against local sources of noise. A particular set of these kind of excitations are the nowadays called Majorana modes. These quasiparticle excitations are identical to their own antiparticles. They inherit their name from the famous physicist Ettore Majorana who proposed a modification
of the Dirac equation in order to describe fermionic particles that are their own antiparticles \cite{Majorana}. 

It has been theorized that Majorana states are formed at the ends of superconductor wires as a consequence of the combined action
of superconductivity, Rashba spin-orbit coupling and Zeeman magnetic effect \cite{Lutchyn,Oreg,Stanescu,Flensberg,Potter,Potter2,Ganga,Egger,Zazunov,Klino1,Klino2,Lim,Lim2,Lim3,Serra}. In a superconductor nanowire electrons play the role of particles while holes of opposite charge and spin perform the role of antiparticles. These kind of systems are modeled as quantum gases with the mentioned interactions and their Majorana modes can only exist at zero energy because antiparticle states have an opposite energy that their particle counterparts. Superconductivity leads to a charge symmetry breaking and allows quasiparticles without a good isospin number to exist. On the other hand, the Rashba effect is a direct result of an inversion asymmetry caused by an electric field in a direction perpendicular to the electron motion while the Zeeman magnetic field breaks the spin rotation symmetry of the system. The combined action of the Rashba and Zeeman effects are needed to create states with a space precessing spin necessary to obtain effective spin less Majorana states at zero energy.

In a finite nanowire Majorana physics necessarily manifests in an approximate way. This is
because of the unavoidable interference between the states on the two opposite ends of the finite 
nanowire.  These states never lie exactly at zero energy and their wave functions always
overlap to some degree.
In long enough nanowires, however, Majorana behavior is seen as energy eigenvalues very close to zero,
protected by a sizable energy gap from the rest of eigenvalues and with exponentially small overlap 
of their wave functions. We can, of course, interpret those near-zero-energy states
as the finite-system Majoranas and consider how their Majorana character is increased or decreased
when some system parameters are varied. That is, we can investigate whether the scenario of protected near-zero-energy
states is better achieved or not after a particular variation.
This approach to Majorana physics is realistic 
since the experiments are always done with finite systems.

Very recently, Mourik and coworkers have reported the detection of such Majoranas in long InSb nanowires ($L=2\, \mu{\rm m}$) \cite{Mourik}. Superconductivity is induced by contact with an NbTiN metallic superconductor caused by the leakage of Cooper pairs into the semiconductor. Superconductivity is maintained all along the few nanometer width of the nanowire,  shorter than the coherence distance of the Cooper pairs. The large g factor of the InSb semiconductor allows the existence of a noticeable Zeeman effect in presence of a moderate magnetic field while at the same time it maintains a strong 
enough spin-orbit interaction. Finally the nanowire is connected to two electrodes, one in each end, and the current is measured. The Majorana mode evidence is a peak at zero voltage in the tunneling differential conductance called zero bias anomaly (ZBA). 
Similar experiments based on the detection of the zero voltage peak have been carried out by different research groups \cite{Das,Churchill,Deng,Fink}.

\begin{figure}
\centering
\resizebox{0.4\textwidth}{!}{%
	\includegraphics{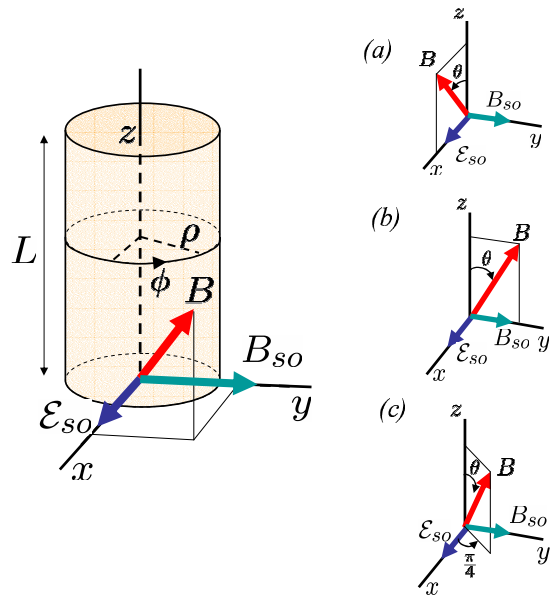}
}
\caption{Hollow semiconductor nanowire of length $L$ with cylindrical geometry. Cylindrical coordinates and unit vectors are
indicated. A magnetic field $\vec{B}$ is applied in an arbitrary direction. 
The orientation of the Rashba electric and magnetic fields, $\vec{\cal E}_{so}$ 
and $\vec{B}_{so}$, are also  indicated.
The  three field tiltings discussed in Sec.\ \ref{sec:2} are indicated 
in (a-c). }
\label{FG20}
\end{figure}

The ZBA only appears in presence of the three required ingredients (superconductivity, Rashba and Zeeman effects) when the system is driven into the topological phase. From the theoretical point of view several authors dealt with one dimensional or two dimensional planar models \cite{Lim,Prada,Alicea}. However, the cylindrical geometry of the nanowire motivated the study made by Lim et al.\ in Ref.\ \cite{Lim3}. In the proposed configuration the spin-orbit electric field ${\cal E}_{so}$ points perpendicularly to the nanowire (see Fig.\ \ref{FG20}) while
the possibility of Majoranas with a radial electric field was discarded in that work 
for any configuration of the remaining parameters.
It is reasonable to expect a weaker total spin-orbit effect if the 
field is radial simply because of the compensation 
for opposite angles. On the other hand, a fixed direction
${\cal E}_{so}$ can originate in the asymmetry induced by the superconductor substrate
on which the nanowire is deposited.

In Ref.\ \cite{Lim3} the effects of magnetic fields pointing into the three Cartesian directions 
were considered, for different values of its magnitude. It was concluded that only a magnetic field pointing along the nanowire axis ($z$) is suited to the creation of the Majoranas, but no discussion about the robustness of this particular configuration was made. The present work addresses this issue, studying the robustness of the Majorana modes to different tiltings of the magnetic field and it also investigates the dependence of the nanowire eigenstates  with other parameters, such as the cylinder radius and the 
spin-orbit strength.

We show that Majorana modes are robust to the tilting of the magnetic field only in a particular direction, while these modes are easily destroyed for tiltings in any other directions. This is relevant to avoid possible non Majorana experimental set-ups, but also as possible procedures to verify the Majorana origin of the ZBA checking its robustness against the theoretical predictions. Remarkably, 
when increasing the nanowire radius and/or the spin-orbit strength we find that more and more Majoranas coexist on the same end, a phenomenon we name 'stacking' of Majoranas.
These coexisting Majoranas  tend to occupy complementary spatial positions on the cylinder edge.
 
In Sec.\ \ref{sec:1} the physical system is introduced while
Sec.\ \ref{sec:2} studies the nanowire spectrum for different tilting
directions of the magnetic field. Section \ref{sec:3} is devoted to study the 
spectrum changes when varying the cylinder radius and the spin-orbit strength. Finally, 
the conclusions can be found in Sec.\ \ref{conclusions}.

\section{Physical model}
\label{sec:1}
The nanowire is modeled as a finite two dimensional semiconductor cylindrical shell with spin-orbit interaction and induced superconductivity, inside an homogeneous Zeeman magnetic field, as shown 
in Fig.\ \ref{FG20}. We assume that due to the superconductive character of the nanowire the quasiparticles move only on the cylinder surface.  Experimentally, this type of tubular systems have been  
produced either as rolled-up quantum wells (e.g.\ in Ref.\ \cite{shaj})
or by epitaxial growth of core/shell nanowires (e.g. in\ Ref.\ \cite{scha}).

The system is described by a Hamiltonian of Bogoliubov-deGennes kind \cite{Lim3},
\begin{equation}
	 \mathcal{H}=\mathcal{H}_{kin}+\mathcal{H}_{R}+\mathcal{H}_Z+\mathcal{H}_{S}\,,
  \label{E30}
\end{equation}
where the successive contributions are kinetic, Rashba spin-orbit, Zeeman and superconductivity term. In this manner, the three ingredients needed to obtain Majorana modes are present in the model: superconductivity, spin-orbit interaction and a Zeeman magnetic field.  The kinetic term depends on the canonical momentum $\vec{\Pi}=\vec{p}+\frac{e}{c}\vec{A}$ where $\vec{p}$ is the usual momentum operator in three dimensions and $\vec{A}$ is the vector potential.  The vector potential models the orbital effects of the magnetic
field on the electron trajectories. Therefore, the kinetic term for electrons is
$\vec{\Pi}^2/2m^* - \mu$,
where $\mu$ is the chemical potential. Since the system has cylindrical symmetry it is convenient to write the Hamiltonian in the cylindrical coordinate system indicated in Fig.\ \ref{FG20}. Being a two dimensional problem the cylindrical coordinates will allow us to express the Hamiltonian as function of two coordinates ($z$ and $\phi$, note that the radius $\rho$ is fixed). Furthermore, the kinetic term can be split into a non-magnetic 
\begin{equation}
	 \mathcal{H}^{(0)}_{kin}=
\left[ \frac{p^2_\phi+p^2_z}{2m^*}-\mu \right]\tau_z\,,
  \label{EQ101}
\end{equation}
and a magnetic term
\begin{eqnarray}
\mathcal{H}^{(1)}_{kin}&=&\frac{\hbar^2}{2m^*}\left[
\frac{\rho}{l^2_z}\frac{p_\phi}{\hbar}\tau_z
+\frac{\rho^2}{4 l^2_z}\tau_z
+2\frac{\rho}{l^2_x}\sin\phi\, \frac{p_z}{\hbar}\tau_z
\right.\nonumber\\
&&\left.
\hspace*{1cm}
-2\frac{\rho}{l^2_y}\cos\phi\, \frac{p_z}{\hbar}
+\frac{\rho^2}{l^4_x}\sin^2\phi\, \tau_z
+ \frac{\rho^2}{l^4_y}\cos^2\phi\, \tau_z 
\right.\nonumber\\
&&\left.
\hspace*{1cm}
-2 \frac{\rho^2 }{l^2_x l^2_y} \sin\phi\cos\phi    \right]\,,
  \label{EQ102}
\end{eqnarray}
where the electron and holes degrees of freedom are represented with the Pauli matrices $\tau_{x,y,z}$ acting on the isospin space and  the magnetic lengths along direction $i\equiv x,y,x$ are $l_i=\sqrt{\hbar c /e B_i}$. Trivially, when a certain magnetic field component vanishes the corresponding magnetic length diverges, giving a vanishing contribution to $\mathcal{H}_{kin}$. 

The Rashba term is obtained assuming the existence of an external electric field 
${\cal E}_{so}$, in direction of the unitary vector  $\vec{u}_{\cal E}$. It is 
proportional to the double product of vectors
$\vec{\sigma}\cdot\left( \vec{\Pi} \times \vec{u}_{\cal E} \right)$,
where $\vec{\sigma}$ represents the vector of Pauli matrices for spin.
If we assume an homogeneous electric field pointing in  $\vec{x}$ direction 
the corresponding Hamiltonian contribution 
can be split, by analogy to the kinetic term, into non magnetic $\mathcal{H}_R^{(0)}$ 
and magnetic $\mathcal{H}_R^{(1)}$ terms as
\begin{eqnarray}
	 \mathcal{H}_{R}^{(0)} &=&
	 \frac{\alpha}{\hbar}\left( p_z\, \sigma_y\tau_z  -  \cos\phi\, p_\phi \, \sigma_z
	 -\frac{i\hbar}{2\rho}\sin\phi\, \sigma_z	 \right)\;, 
  \label{E40}\\
	 \mathcal{H}_{R}^{(1)} &=& 
\alpha\rho\left(
\frac{\sin\phi}{l_x^2}\sigma_y\tau_z
-\frac{\cos\phi}{l_y^2}\sigma_y
-\frac{\cos\phi}{2l_z^2}\sigma_z
\right)\; .
  \label{E41}
\end{eqnarray}

As mentioned in Sec.\ \ref{intro} another natural choice for the Rashba term
could be a radial field 
($\vec{u}_{\cal E}=\vec{u}_\rho$) but, in agreement with Ref.\ \cite{Lim3},
we have checked that this does not lead 
to the formation of any
Majorana-like states. 
For a long-enough cylinder in absence of magnetic field one expects the first contribution to Eq.\ (\ref{E40}) dominate
the others for large values of $p_z$, thus leading to the usual interpretation of the Rashba term as an effective 
magnetic field $\vec{B}_{so}\propto \alpha p_z\vec{y}$, in $y$ direction and thus coupling with $\sigma_y$. 

The Zeeman term for an external magnetic field along $\vec{n}$ reads
\begin{equation}
	 \mathcal{H}_Z = \Delta_B\, \vec{\sigma}\cdot \vec{n}\,.
  \label{EQ105}
\end{equation}
Finally, the superconductor contribution  $\mathcal{H}_{S}$ is 
\begin{equation}
	 \mathcal{H}_S = \Delta_0 \tau_x \,,
  \label{EQ106}
\end{equation}
where $\Delta_0$ is the Cooper-pair breaking energy. The superconduction pairing term couples  
opposite isospin states and
arises from a mean field approximation over electron interactions shielded by the atomic network.

Notice that the Zeeman effect allows the closing of the superconductor energy gap, shifting some quasiparticle states to zero energy while the Rashba spin-orbit term interaction introduces anti crossings in the system spectrum that separate the Majorana zero energy states from the others. The Rashba term arises from the self interaction between an electron (or hole) spin with its own motion due to the presence of a transverse electric field ${\cal E}_{so}$, perceived as an effective magnetic field in the rest frame of the quasiparticle. 
Spin-orbit effects in cylindrical shells 
similar to the ones considered here but without superconductivity
have been studied in Refs.\ \cite{trush,bring}.

The Rashba spin-orbit and Zeeman effects depend on the parameters $\alpha$ and $\Delta_B$, respectively. Since we consider a nanowire made of an homogeneous material inside a constant magnetic field these parameters are assumed to be homogeneous.
 The potential, not included in the Hamiltonian, is taken as zero inside the nanowire and infinite outside. Therefore it is included into the calculations as a boundary condition. Summarizing, the Hamiltonian depends on the superconducting gap $\Delta_0$, the Zeeman energy $\Delta_B$ and the Rashba coupling strength $\alpha$, as well as  on the direction of the magnetic field $\vec{n}$.

The explicit energy contributions from the orbital effects
of the magnetic field are contained in the kinetic ${\cal H}_{\it kin}^{(1)}$
and ${\cal H}_R^{(1)}$ terms. Below we will  also study the results when these 
contributions are omitted, finding in general that orbital effects are very important and can not be neglected. The orbital terms increase with
the cylinder radius due to contributions that are linear and quadratic in $\rho$ 
and they contain the 
dependence on curvature of the model.

To obtain the energy spectrum the total Hamiltonian is diagonalized numerically using as a basis states 
characterized as 
\begin{equation}
|n m s_\sigma s_\tau\rangle
\qquad
\left\{
\begin{array}{rcl}
n&=&1,2,\dots\; ,\\
m&=&0,\pm1,\pm2,\dots\; ,\\
s_\sigma&=& \pm\; ,\\
s_\tau&=&\pm\; ,
\end{array}
\right.
\end{equation}
where $n$ is the quantum number associated with the infinite square well eigenstates ($z$ direction), 
$m$ is the $L_z$ eigenvalue (angular momentum along $z$) while $s_\sigma$ and $s_\tau$ are the spin and isospin quantum numbers. The numerical algorithm works in dimensionless units, defining energy and length units as
\begin{eqnarray}
	 E_U &\equiv& \frac{\hbar^2}{m^* \rho^2} \,,
  \label{EQ107}\\
	 L_U&\equiv& \rho \,.
  \label{EQ108}
\end{eqnarray}

More specifically, assuming a cylinder radius of 
$\rho=50$ nm, and $m^*=0.015m_e$ for InSb,
one has $E_U=2.03$ meV and $L_U=50$ nm.
From Ref.\ \cite{Mourik}, typical values for the Rashba parameter and the superconducting gap
are $\alpha\approx 20$ meVnm and $\Delta_0\approx 0.25$ meV, which in scaled
units are $\alpha\approx 0.2 E_UL_U$ and $\Delta_0\approx 0.12 E_U$.
We will present below the results in the generalized units,
assuming an arbitrary $m^*$ and $\rho$, 
since the conversion for each material and cylinder radius is just 
a trivial scaling.

\begin{figure} 
\centering
\resizebox{0.5\textwidth}{!}{%
	\includegraphics{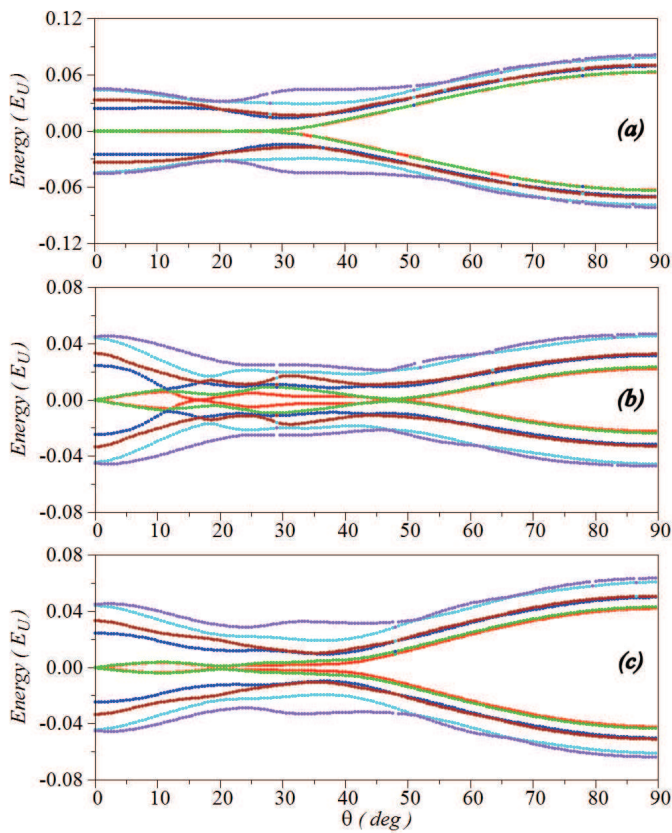}
}
\caption{Eigenenergies (including orbital effects) as a function of the 
magnetic tilting angle from the $z$ axis, for different orientations: 
a) $y=0$ plane, b) $x=0$ plane,
 c)  $x=y$ plane.
 The parameters used are  $\Delta_B=0.2 E_U$, $\Delta_0=0.12E_U$, $\alpha=0.2E_U L_U$, $\mu=0$ and $L_z=30L_U$. The number of basis states are $N_n=51$ and $N_m=31$. Only the lowest 6 pairs of eigenvalues are displayed.}
\label{FG23}
\end{figure}

\section{Results on tilted fields}
\label{sec:2}
In Ref. \cite{Lim3} it was shown that the critical fields corresponding to transitions into the topological phase of the nanowire were strongly dependent on the external magnetic field orientation. Considering a nanowire whose dominant spin-orbit effective magnetic field $\vec{B}_{so}$ points along $\vec{y}$ 
(Fig.\ \ref{FG20}), if the external magnetic field $\vec{B}$ points also into this same direction then the Rashba $\vec{B}_{so}$ contribution and the Zeeman term commute and no anti crossings are induced in the spectrum. 
As a consequence, no spin less zero energy modes are observed since the Majorana conditions are not met. 
It was also shown 
in Ref.\ \cite{Lim3} that, due to the strong influence of the orbital effects,
Majorana modes are neither possible with an external magnetic field along $\vec{x}$. 
Therefore, for the three main Cartesian orientations Majorana modes are found only when the magnetic field is pointing in the $\vec{z}$ direction, and for strong-enough values.
We address in this section the study of the 
transition between these configurations and, particularly, the robustness of the Majoranas 
(or their lack of it) due to fluctuations in the magnetic field direction. 

Figures \ref{FG23}a-c show how the spectrum of a cylindrical nanowire 
changes when the magnetic field is tilted from the $z$ axis in the 
directions sketched
in Fig.\ \ref{FG20},
while its magnitude is maintained.
Notice that Fig \ref{FG23}a proves that the Majorana modes are robust to a tilting of the magnetic field from $\vec{z}$ towards $\vec{x}$, up to almost $30^\circ$ when the tilting is 
done within the $y=0$ plane. On the contrary, 
Fig.\ \ref{FG23}b shows that the Majorana modes break almost immediately 
if the tilting is done towards $\vec{y}$ in the $x=0$ plane. 
Zero energy crossings are found in Fig.\ \ref{FG23}b
for $\theta\approx15^\circ$ and, when the magnetic field is further tilted, also around $\theta\approx50^\circ$, but this kind of crossings can be labeled as accidental since they only occur at specific 
points.

 Majorana behavior is characterized by eigenvalues lying very close to zero for a continuous range of parameter values, protected from nearby states by a sizable energy gap much larger that their own energy. 
Accidental zero energy crossings, on the contrary,
occur at specific values and in this sense they 
are weak against any small parameter fluctuation. 
Notice that Fig.\ \ref{FG23} shows that
 tilting directions towards $\vec{x}$ and $\vec{y}$ are not equivalent,
 in spite of being perpendicular between them and 
also with $\vec{z}$. As mentioned above, differences are 
due to the relative orientation with respect to the effective Rashba field $\vec{B}_{so}$, 
that in our case points in the $\vec{y}$ direction. 

\begin{figure} 
\centering
\resizebox{0.5\textwidth}{!}{%
	\includegraphics{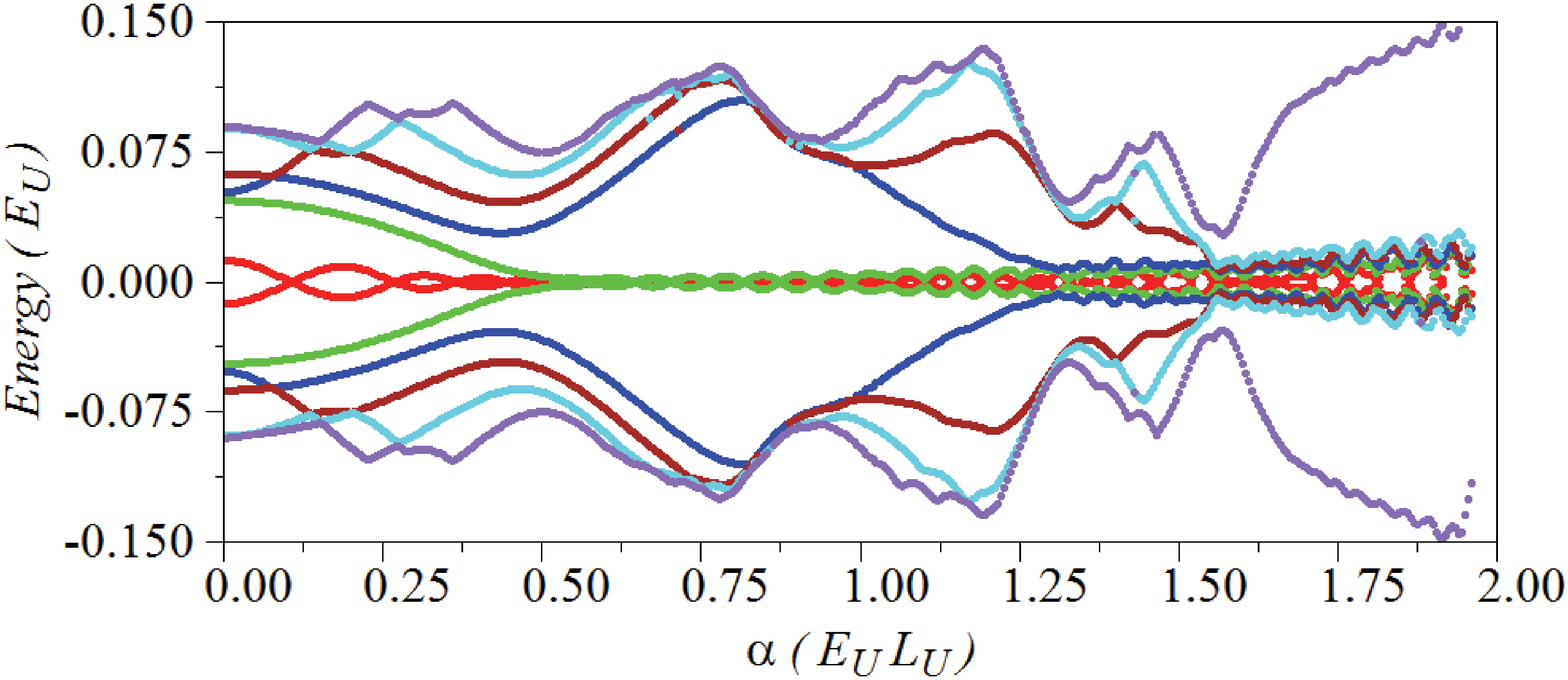}
}
\caption{Eigenenergies of a cylinder (considering the orbital effects) as a function of spin-orbit coupling. The rest of the parameters are the same of Fig.\ \ref{FG23}, including the number of basis states and the number of displayed eigenvalues.}
\label{FG24}
\end{figure}

\begin{figure} 
\centering
\resizebox{0.5\textwidth}{!}{%
	\includegraphics{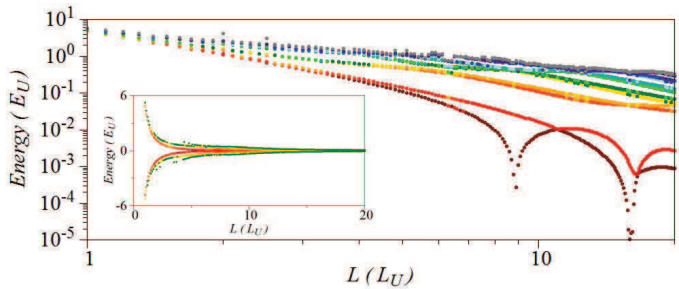}
}
\caption{Dependence of the spectrum of eigenergies of a cylinder with respect to its length. The rest of the parameters are the same of  Fig. \ref{FG23}.
The main plot shows only positive energy eigenvalues in logarithmic scales
while the inset contains both positive and negative eigenvalues in linear 
scales.
}
\label{FG25}
\end{figure}

On the other hand, if the magnetic field is tilted away from $\vec{z}$ in the $x=y$ plane (Fig.\ \ref{FG23}c)
the resulting spectrum has mixed features with respect to those of Figs. \ref{FG23}a and \ref{FG23}b. 
The Majoranas break up for small angles but the resulting fermionic modes remain close to zero energy up to almost $40^\circ$. Therefore, cylinder Majoranas are robust to deviations of the external magnetic field from the cylinder axis only within the plane perpendicular to the spin-orbit effective magnetic field $\vec{B}_{so}$.

\section{Dependence on nanowire characteristics}
\label{sec:3}
\subsection{Spin-orbit coupling}
Figure \ref{FG24} shows a spectrum for a cylindrical nanowire with the same parameters used in Fig.\ \ref{FG23} as a function of the spin-orbit coupling strength $\alpha$. The magnitude and the direction of the magnetic field are kept constant to $\Delta_B=0.2 E_U$ and $\vec{z}$, respectively. In principle, this change can be experimentally realized controlling the intensity of the transverse electrical field
$\vec{\cal E}_{so}$. As observed in Fig.\ \ref{FG24}, for zero spin-orbit the states lying closer to zero 
are two states of fermionic type at finite energies.  Increasing the value of $\alpha$ those
two states evolve into a Majorana pair as their energy approaches zero.  
Note also that additional Majorana modes become activated sequentially, in more or less regular intervals of the coupling constant $\alpha$. Each time a new Majorana mode arises, the protection (energy gap) diminishes for a short range of values, to increase again once the Majorana is fully formed. 

Since the spin-orbit coupling is also affected by the orbital terms, when high values of $\alpha$ are achieved the spectrum of the modes near zero energy 
in Fig.\ \ref{FG24}
becomes more and more oscillating, blurring away the Majorana character of these modes. Although the process of Majorana activation still works for the higher values of spin-orbit coupling, these states do not
become Majorana-like due to their 
sizable energies that 
keep oscillating with increasing amplitudes. Figure \ref{FG24} suggests an optimal range for the formation and coexistence of Majoranas
when the amplitude of the energy oscillation around zero attains its minimum,
with two pairs coexisting for $ \alpha\approx 0.6 E_U L_U $.

\begin{figure} 
\centering
\resizebox{0.5\textwidth}{!}{%
\includegraphics{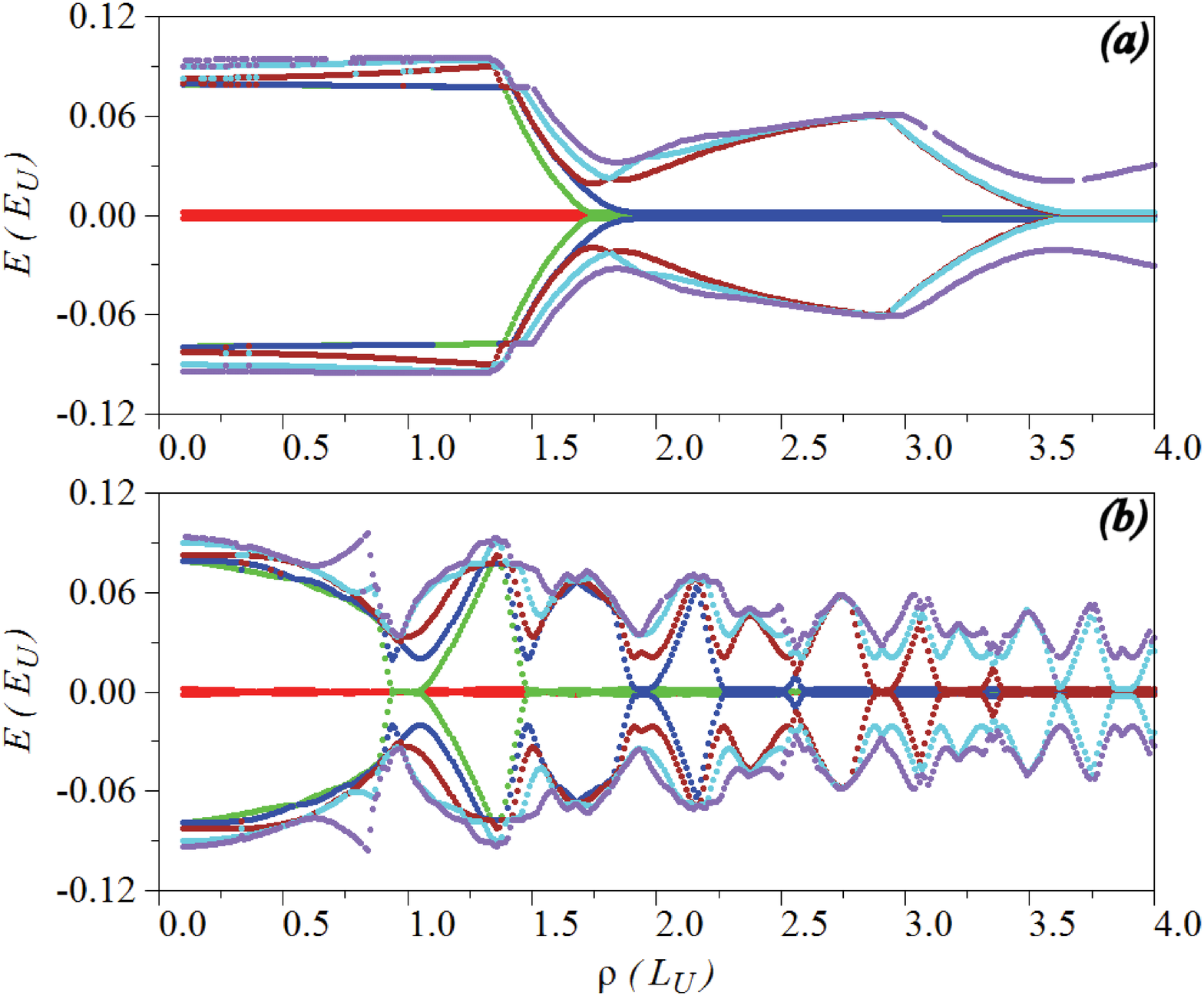}
}
\caption{Dependence of the spectrum of eigenergies of a cylinder with respect to its radius. The rest of the parameters are the same as in Fig. \ref{FG23}. 
Panel (a) is the result obtained when neglecting the orbital contributions
of the magnetic field  ${\cal H}_{kin}^{(1)}$
and ${\cal H}_R^{(1)}$, while (b) is for the complete Hamiltonian.
Note that the length unit $L_U$ in this figure is taken as an arbitrary reference distance.
}
\label{FG26}
\end{figure}

\subsection{Cylinder geometry}
The length and radius of the nanowire are not tunable parameters in the same sense as the external magnetic and electric fields, but can be controlled by fabricating different versions of the nanowire with different dimensions. As can be seen in Fig.\ \ref{FG25} Majoranas are not possible for very short nanowires because 
both ends greatly interfere with each other creating a fermionic state. Qualitatively, for the particular choice of parameters in this figure the nanowire length must exceed at least six times its radius in order to be able to hold a Majorana
state with energy lower than, roughly, $0.01 E_U$. 
When displayed in logarithmic scales,
Fig.\ \ref{FG25} also shows that
the energy of the states lying closer to zero oscillate as a function of
the cylinder radius, similarly to the $\alpha$ dependence of Fig.\ \ref{FG24}.

On the other hand, increasing the radius of the nanowire leads to a stacking of Majoranas (Fig.\ \ref{FG26}).
A qualitatively similar coalescence is found for Shockley states at the surface
of finite crystals \cite{Shock}, which is suggesting
a generic behavior for edge states.
A technical point in the analysis of a varying cylinder radius $\rho$ is that this can no
longer be identical to the length unit $L_U$.
To avoid ambiguities, we take an arbitrary length as a reference $L_U$ and measure all distances (including the varying cylinder radius in Fig.\ \ref{FG26}) with respect to this unit. 

As shown in Fig.\ \ref{FG26}a, if the orbital effects are disregarded, the zero point energy contribution of the angular momentum decreases as the radius becomes larger, allowing the activation of the Majorana modes, creating this way a stacking of zero modes at regular intervals of the radius. Each Majorana pair
is associated with an angular momentum quantum number $m$. Since $m$ and $-m$ eigenenergies are almost degenerate, two pairs are activated at around the same value of the radius. As they are not completely degenerate by the action of the transversal electric field, there is a slight difference in their activation value. 

With orbital effects the spectrum becomes more involved (see Fig.\ \ref{FG26}b). First, the Majorana pairs are activated one by one as the value of the radius increases. Furthermore, each pair is destroyed shortly after its activation, developing into a fermion pair of states above the gap, and is re-formed again for larger values. This behavior is due to the increase with $\rho$ of the orbital 
terms ${\cal H}_{kin}^{(1)}$ and ${\cal H}_R^{(1)}$.
There is a competition between the different effects of the spin-orbit terms, some of them helping to create the Majorana and some of them trying to destroy the Majorana, with the former ones eventually wining for high values of the radius. Nevertheless, we stress the general tendency of Majorana mode stacking at zero energy for high values of the radius.

\begin{figure*} 
\centering
\vspace*{5cm}
\resizebox{0.9\textwidth}{!}{%
	\includegraphics{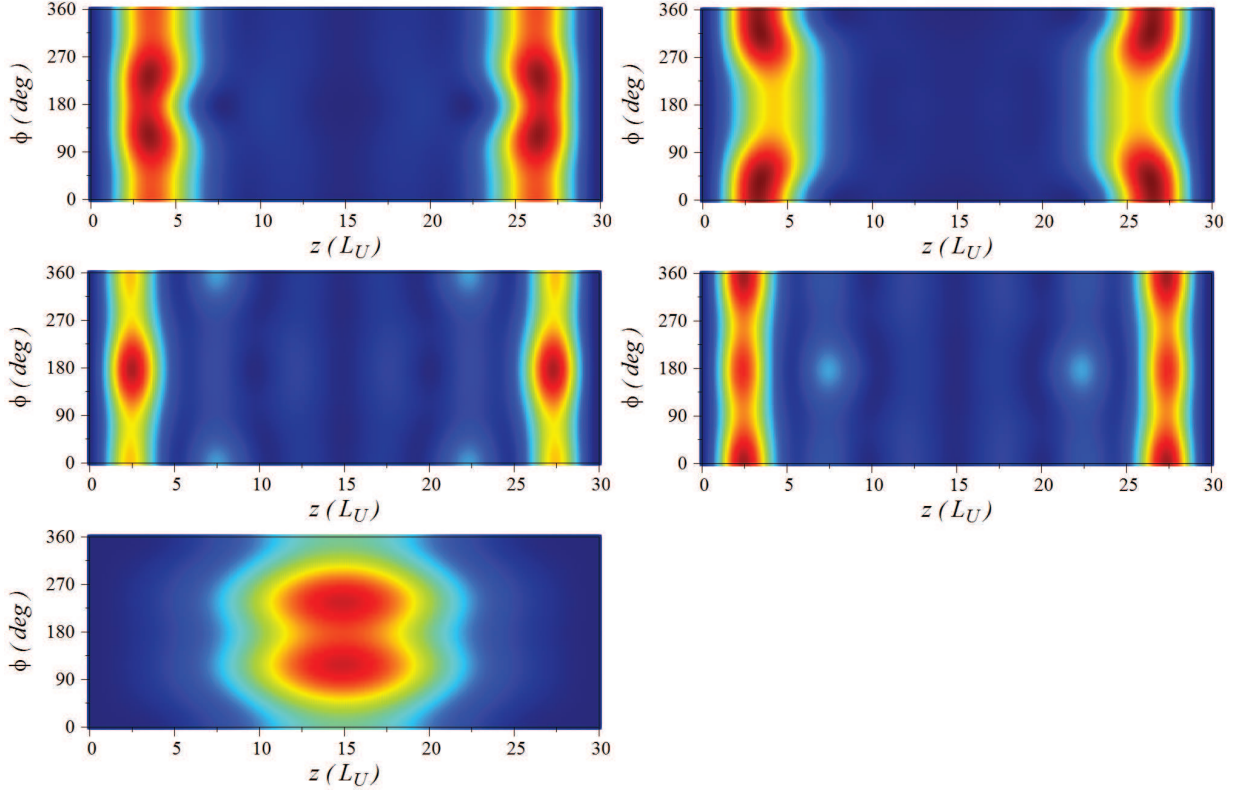}
}
\caption{Density distributions of the lower cylinder eigenmodes. The parameters used are $\Delta_B=0.2 E_U$, $\Delta_0=0.12E_U$, $\alpha=0.2E_U L_U$, $\mu=0$, $\rho=3.2L_U$ and $L_z=30L_U$. The densities of the first positive energy solutions are plotted. All of them are Majorana modes with the exception of the panel at the bottom that is a fermionic one.}
\label{FG27}
\end{figure*}

We finally discuss the spatial distribution of the probability densities 
associated with the Majorana-like states. A clear signature of  
Majorana character is a strong localization at the nanowire edges, with
very small central overlaps. Figure \ref{FG27} shows in a particular example 
that this is indeed observed, with a very clear difference between 
Majorana-like and fermionic modes. 
Figure \ref{FG27} also shows that the above mentioned stacking is possible for high values of the radius 
because there is room for orthogonal wave functions, with almost non overlapping density distributions, to be formed on the same nanowire end. The higher the radius the larger the suitable region for the Majoranas to be formed, where more and more orthogonal zero energy wave functions can accommodate
in complementary regions.

\section{Conclusions}
\label{conclusions}
The diagonalization of the Hamiltonian for a two dimensional cylindrical shell has allowed us 
to discuss the eigenenergies and eigenstates of a finite nanowire. In particular, 
we have focussed on the Majorana state wave functions and the parameter configurations leading to their appearance. 
The resilience of the Majorana states to the tilting of the magnetic field, as far as $30^\circ$ into the 
$\vec{\cal E}_{so}$ direction, has been shown; as well as the lack of it if one component points into the $\vec{B}_{so}$ direction. We have also learned how a strong spin-orbit coupling is needed in order to have Majoranas but, at the same time, how orbital terms affect the role of the spin-orbit coupling. This means that for too high values of the spin-orbit coupling Majorana modes get blurred due to orbital effects. 

Our main result is the stacking effect of Majorana modes for 
increasing nanowire radius and 
high enough values of the magnetic field. This is a novelty in comparison with one dimensional models,  and it is an exclusive property of two dimensional models. It means that we can tune the number of localized Majorana modes on each nanowire end. Although not explored in this work, this is hinting
interesting transport properties that could be useful to experimentally confirm the presence
of Majoranas  in nanowires and it will be matter of future work.

\section*{Acknowledgments}
Work supported by MINECO Grant No.\ FIS2011-23526,
the Conselleria d'Educaci\'o, Cultura i Universitats
(CAIB) and FEDER. We hereby acknowledge the PhD grant provided by the University 
of the Balearic Islands.

 \bibliographystyle{epjc}
 \bibliography{Articulo_2EPJ}

\end{document}